\documentclass[showpacs,preprintnumbers,twocolumn,
amsmath,amssymb,groupedaddress,superscriptaddress]{revtex4-1}

\usepackage{graphicx}
\usepackage{dcolumn}
\usepackage{bm}

\begin{document}

\title{Symmetry energy of deformed neutron-rich nuclei}

\author{M.K. Gaidarov}
\affiliation{Institute for Nuclear Research and Nuclear Energy,
Bulgarian Academy of Sciences, Sofia 1784, Bulgaria}

\author{A.N. Antonov}
\affiliation{Institute for Nuclear Research and Nuclear Energy,
Bulgarian Academy of Sciences, Sofia 1784, Bulgaria}

\author{P. Sarriguren}
\affiliation{Instituto de Estructura de la Materia, IEM-CSIC,
Serrano 123, E-28006 Madrid, Spain}

\author{E. Moya de Guerra}
\affiliation{Departamento de Fisica Atomica, Molecular y Nuclear,
Facultad de Ciencias Fisicas, Universidad Complutense de Madrid,
E-28040 Madrid, Spain}

%\date{\today}

\begin{abstract}
The symmetry energy, the neutron pressure and the asymmetric
compressibility of deformed neutron-rich even-even nuclei are
calculated on the examples of Kr and Sm isotopes within the
coherent density fluctuation model using the symmetry energy as a
function of density within the Brueckner energy-density
functional. The correlation between the thickness of the neutron
skin and the characteristics related with the density dependence
of the nuclear symmetry energy is investigated for isotopic chains
of these nuclei in the framework of the self-consistent
Skyrme-Hartree-Fock plus BCS method. Results for an extended chain
of Pb isotopes are also presented. A remarkable difference is
found in the trend followed by the different isotopic chains: the
studied correlations reveal a smoother behavior in the Pb case
than in the other cases. We also notice that the neutron skin
thickness obtained for $^{208}$Pb with SLy4 force is found to be
in a good agreement with recent data.
\end{abstract}


\maketitle

\section{Introduction}

The study of the nuclear matter symmetry energy that essentially
characterizes the isospin-dependent part of the equation of state
(EOS) of asymmetric nuclear matter (ANM) is currently an exciting
topic of research in nuclear physics
\cite{Danielewicz2002,Famiano2006,Shetty2007,Centelles2009}. In
fact, applications of ANM are broad, ranging from the structure of
rare isotopes \cite{Brown2000,Typel2001} to the properties of
neutron stars \cite{Horowitz2001,Lattimer2007} and the dynamical
process of nuclear reactions \cite{Baran2005}. The transition from
ANM to finite nuclei is a natural and important way to learn more
about the nuclear symmetry energy which is poorly constrained by
experimental data on ground-state nuclear properties.

The ground states of atomic nuclei are characterized by different
equilibrium configurations related to corresponding geometrical
shapes. The study of the latter, as well as the transition regions
between them, has been a subject of a large number of theoretical
and experimental studies (for a review, see, for example,
Ref.~\cite{Wood92} and references therein). The position of the
neutron drip line is closely related to the neutron excess and the
deformation in nuclei. Deformed nuclei are expected in several
regions near the neutron drip line
\cite{Stoitsov2003,Goriely2009}. In some cases, the deformation
energy can impact their existence. For instance, it has been
predicted that there exist particle-bound even-even nuclei that
have, at the same time, negative two-neutron separation energies
caused by shape coexistence effects \cite{Stoitsov2003}. In fact,
the nuclear deformation increases the surface area, thus leading
to a larger surface symmetry energy in a neutron-rich nucleus with
a deformed shape. Conversely, the precise determination of the
surface symmetry energy is important to describe the deformability
of neutron-rich systems and also to validate theoretical
extrapolations.

The theoretical treatment of the ground-state properties of
deformed nuclei is usually made in the framework of mean-field
approaches, for instance, the nonrelativistic self-consistent
Hartree-Fock+BCS approach with density-dependent Skyrme
interaction
\cite{Vautherin73,Bender2003,Guzman2007,Sarriguren2007,Gaidarov2009},
or the Hartree-Fock-Bogoliubov (HFB) approximation that uses, in
particular, the Gogny force containing a finite-range interaction
\cite{Decharge80,Robledo2008}. Also the relativistic mean-field
(RMF) models with different types of parameter sets have been
widely used. For instance, the RMF theory with parameter set
FSUGold has been recently employed to study the binding energies,
quadrupole deformations, charge radii and neutron skins of
rare-earth even-even nuclei ranged from $Z=58$ to $Z=70$ and some
deformed nuclei (Nd, Sm, Gd, Dy) in other regions
\cite{Sheng2010a}. This new parameter set that includes the
nonlinear coupling between the isoscalar and isovector mesons is
shown to reproduce successfully the ground-state properties of
deformed nuclei. Also, the same parameter set FSUGold can
successfully reflect the shell effect of the neutron magic number
$N=82$ thus leading to a good description of the ground-state
properties of Sn, Te, Xe, and Ba isotopes \cite{Sheng2010b}.

Nowadays, the experimental information about the symmetry energy
is fairly limited. The need to have information for this quantity
in finite nuclei, even theoretically obtained, is a major issue
because it allows one to constrain the bulk and surface properties
of the nuclear energy-density functionals (EDFs) quite
effectively. For example, the traditional Skyrme EDF based on the
leptodermous expansion of the smooth nuclear energy was used in
Ref.~\cite{Nikolov2011}, where the strongly correlated symmetry
and surface symmetry terms in this expansion are resolved by
considering data on deformed neutron-rich nuclei in which the
surface symmetry term is amplified. By starting from microscopic
nucleon self-energies in nuclear matter, the authors of
Ref.~\cite{Niksic2008} have made a comprehensive study of
deformation properties of relativistic nuclear EDFs for a set of
64 axially deformed nuclei in the mass regions $A\approx $150--180
and $A\approx $230--250.

In our recent work \cite{Gaidarov2011} the Brueckner EDF for
infinite nuclear matter was applied to calculate nuclear
quantities of medium-heavy and heavy Ni, Sn, and Pb nuclei that
include surface effects, namely the nuclear symmetry energy $s$,
the neutron pressure $p_{0}$, and the asymmetric compressibility
$\Delta K$. For this purpose, a theoretical approach that combines
the deformed HF+BCS method with Skyrme-type density-dependent
effective interactions \cite{Vautherin73} and the coherent density
fluctuation model (CDFM) \cite{Ant80,AHP} was used. We would like
to note the capability of the CDFM to be applied as an alternative
way to make a transition from the properties of nuclear matter to
the properties of finite nuclei. We have found that there exists
an approximate linear correlation between the neutron skin
thickness $\Delta R$ of even-even nuclei from the Ni ($A=74-84$),
Sn ($A=124-152$), and Pb ($A=206-214$) isotopic chains and their
nuclear symmetry energies. A similar linear correlation between
$\Delta R$ and $p_{0}$ was also found to exist, while the relation
between $\Delta R$ and $\Delta K$ turned out to be less
pronounced. The kinks displayed by Ni and Sn isotopes and the lack
of such kink in the Pb chain considered \cite{Gaidarov2011} were
shown to be mainly due to the shell structure of these exotic
nuclei but they deserve further analysis within the used
theoretical approach.

Another interesting question is to explore how the nuclear
symmetry energy changes in the presence of deformation and
correlates with the neutron skin thickness within a given isotopic
chain. In Ref.~\cite{Sarriguren2007}, the effects of deformation
on the skin formation were studied in Kr isotopes that are
well-deformed nuclei. It has been shown from the analysis on
$^{98,100}$Kr nuclei that although the profiles of the proton and
neutron densities, as well as the spatial extensions change with
the direction in both oblate and prolate shapes, the neutron skin
thickness remains almost equal along the different directions
perpendicular to the surface. Thus, a very weak dependence of the
neutron skin formation on the character of the deformation was
found \cite{Sarriguren2007}.

In the present work an investigation of possible relation between
the neutron skin thickness and the basic nuclear matter properties
in deformed finite nuclei, such as the symmetry energy at the
saturation point, symmetry pressure, and asymmetric
compressibility, is carried out for chains of deformed
neutron-rich even-even Kr ($A=82-96$) (including, as well, the
case of some extreme neutron-rich nuclei up to $^{120}$Kr) and Sm
($A=140-156$) isotopes, following the theoretical method of
Ref.~\cite{Gaidarov2011}. We also present for comparison results
for an extended chain of Pb ($A=202-214$) isotopes. This is
motivated by the significant interest (in both experiment
\cite{prex,Abrahamyan2012,Tamii2011} and theory
\cite{Moreno2010,Horowitz2012,Roca2011,Piekarewicz2012}) to study
the neutron distribution and rms radius in $^{208}$Pb, aiming at
precise determinations of the neutron skin in this nucleus. In
addition to the interest that this study may have by itself as
well as in combination with the previous calculations of
Ref.~\cite{Gaidarov2011}, we give some numerical arguments in
proof of the existence of kinks in Ni and Sn isotopic chains that
are not present in the Pb chain. The kinks are produced because of
the sensitivity of the symmetry energy and neutron pressure to the
shell structure (see, for instance, the discussion in
Refs.~\cite{Gaidarov2011,Warda2010,Mekjian2012}).

The paper is organized as follows. In Sec. II we present a brief description of
the theoretical formalism (definitions of ANM properties, CDFM basic expressions,
Brueckner energy-density functional, Hartree-Fock+BCS densities) used to unveil a
possible correlation between the neutron skin thickness and the nuclear matter
characteristics of the considered isotopic chains. Section III contains our results
with a discussion on the obtained relationships and on the presence of kinks.
The concluding remarks are drawn in Sec. IV.

\section{Theoretical framework}

We study in the present work the symmetry energy $s(\rho)$ and
related quantities of finite deformed nuclei on the basis of the
corresponding definitions for ANM. The quantity $s^{ANM}(\rho)$,
which refers to the infinite system and therefore neglects surface
effects, is related to the second derivative of the energy per
particle $E(\rho,\delta)$ using its Taylor series expansion in
terms of the isospin asymmetry $\delta=(\rho_{n}-\rho_{p})/\rho$
($\rho$, $\rho_{n}$ and $\rho_{p}$ being the baryon, neutron and
proton densities, respectively) (see, e.g.,
\cite{Gaidarov2011,Diep2003,Chen2011}):
%
%\begin{eqnarray}
%s^{ANM}(\rho)&=&\frac{1}{2}\left.
%\frac{\partial^{2}E(\rho,\delta)}{\partial\delta^{2}} \right
%|_{\delta=0} \nonumber \\ &=& \left.
%a_{4}+\frac{p_{0}^{ANM}}{\rho_{0}^{2}}(\rho-\rho_{0})+\frac{\Delta
%K^{ANM}}{18\rho_{0}^{2}}(\rho-\rho_{0})^{2}+ \right.
%\cdot\cdot\cdot \;.
%\label{eq:1}
%\end{eqnarray}
%
%
\begin{widetext}
\begin{equation}
s^{ANM}(\rho)=\frac{1}{2}\left.
\frac{\partial^{2}E(\rho,\delta)}{\partial\delta^{2}} \right
|_{\delta=0} = \left.
a_{4}+\frac{p_{0}^{ANM}}{\rho_{0}^{2}}(\rho-\rho_{0})+\frac{\Delta
K^{ANM}}{18\rho_{0}^{2}}(\rho-\rho_{0})^{2}+ \right.
\cdot\cdot\cdot \;.
\label{eq:1}
\end{equation}
\end{widetext}
In Eq.~(\ref{eq:1}) the parameter $a_{4}$ is the symmetry energy
at equilibrium ($\rho=\rho_{0}$). In ANM the pressure
$p_{0}^{ANM}$ and the curvature $\Delta K^{ANM}$ are:
\begin{equation}
p_{0}^{ANM}=\rho_{0}^{2}\left.
\frac{\partial{s^{ANM}(\rho)}}{\partial{\rho}} \right
|_{\rho=\rho_{0}}, \label{eq:2}
\end{equation}
\begin{equation}
\Delta K^{ANM}=9\rho_{0}^{2}\left.
\frac{\partial^{2}s^{ANM}(\rho)}{\partial\rho^{2}} \right
|_{\rho=\rho_{0}}. \label{eq:3}
\end{equation}
The "slope" parameter $L^{ANM}$ is defined as
\begin{equation}
L^{ANM}=\frac{3p_{0}^{ANM}}{\rho_{0}}.
\label{eq:4}
\end{equation}
In general, the predictions for the symmetry energy vary quite
substantially: e.g., $a_{4}\equiv s(\rho_{0})=28-38$ MeV while an
empirical value of $a_{4}\approx 29$ MeV has been extracted from
finite nuclei by fitting the ground-state energies using the
generalized Weizs\"{a}cker mass formula (see, e.g.,
Ref.~\cite{Danielewicz}). By using the experimental pygmy
strength, an average value of $a_{4}=32.0\pm1.8$ MeV was obtained
from the $^{130,132}$Sn analysis \cite{Klimkiewicz2007}, which is
within the acceptable range of values of $a_{4}$ to be around 32.5
MeV coming from various experiments using different experimental
probes (for a recent status, see, for example,
Ref.~\cite{Tsang2012} and references therein).

In Ref.~\cite{Gaidarov2011} we calculated the symmetry energy, the
pressure and slope, as well as the curvature for {\it finite}
nuclei applying the coherent density fluctuation model (suggested
and developed in Refs.~\cite{Ant80,AHP}). In the CDFM the one-body
density matrix $\rho({\bf r},{\bf r^{\prime}})$ of the nucleus is
written as a coherent superposition of the one-body density
matrices $\rho_{x}({\bf r},{\bf r^{\prime}})$ for spherical
"pieces" of nuclear matter called "fluctons" with densities
$\rho_{x}({\bf r})=\rho_{0}(x)\Theta(x-|{\bf r}|)$,
$\rho_{0}(x)=3A/4\pi x^{3}$:
\begin{equation}
\rho({\bf r},{\bf r^{\prime}})=\int_{0}^{\infty}dx |f(x)|^{2}
\rho_{x}({\bf r},{\bf r^{\prime}})
\label{eq:5}
\end{equation}
with
\begin{eqnarray}
\rho_{x}({\bf r},{\bf r^{\prime}})&=&3\rho_{0}(x)
\frac{j_{1}(k_{F}(x)|{\bf r}-{\bf r^{\prime}}|)}{(k_{F}(x)|{\bf
r}-{\bf r^{\prime}}|)}\nonumber \\  & \times & \Theta \left
(x-\frac{|{\bf r}+{\bf r^{\prime}}|}{2}\right ),
\label{eq:6}
\end{eqnarray}
where $j_{1}$ is the first-order spherical Bessel function,
\begin{equation}
k_{F}(x)=\left(\frac{3\pi^{2}}{2}\rho_{0}(x)\right )^{1/3}\equiv
\frac{\alpha}{x}
\label{eq:7}
\end{equation}
with
\begin{equation}
\alpha=\left(\frac{9\pi A}{8}\right )^{1/3}\simeq 1.52A^{1/3}
\label{eq:8}
\end{equation}
is the Fermi momentum of the nucleons in the "flucton" with a
radius $x$. In Eq.~(\ref{eq:5}) $|f(x)|^{2}$ is the weight
function that in the case of monotonically decreasing local
densities ($d\rho(r)/dr\leq 0$) can be obtained using a known
density distribution for a given nucleus:
\begin{equation}
|f(x)|^{2}=-\frac{1}{\rho_{0}(x)} \left. \frac{d\rho(r)}{dr}\right
|_{r=x}
\label{eq:9}
\end{equation}
with the normalization $\int_{0}^{\infty}dx |f(x)|^{2}=1$.

The main assumption of the CDFM is that properties of {\it finite
nuclei} can be calculated using the corresponding ones for nuclear
matter, folding them with the weight function $|f(x)|^{2}$. Along
this line, in the CDFM the symmetry energy for finite nuclei and
related quantities are assumed to be infinite superpositions of
the corresponding ANM quantities weighted by $|f(x)|^{2}$:
\begin{equation}
s=\int_{0}^{\infty}dx|f(x)|^{2}s^{ANM}(x),
\label{eq:10}
\end{equation}
\begin{equation}
p_{0}=\int_{0}^{\infty}dx|f(x)|^{2}p_{0}^{ANM}(x),
\label{eq:11}
\end{equation}
\begin{equation}
\Delta K=\int_{0}^{\infty}dx|f(x)|^{2}\Delta K^{ANM}(x).
\label{eq:12}
\end{equation}
The explicit forms of the ANM quantities $s^{ANM}(x)$,
$p_{0}^{ANM}(x)$, and $\Delta K^{ANM}(x)$ in Eqs.~(\ref{eq:10}),
(\ref{eq:11}), and (\ref{eq:12}) are defined below. They have to
be determined within a chosen method for description of the ANM
characteristics. In the present work, as well as in
Ref.~\cite{Gaidarov2011}, considering the pieces of nuclear matter
with density $\rho_{0}(x)$, we use for the matrix element $V(x)$
of the nuclear Hamiltonian the corresponding ANM energy from the
method of Brueckner {\it et al.} \cite{Brueckner68,Brueckner69}:
\begin{equation}
V(x)=A V_{0}(x)+V_{C}-V_{CO},
\label{eq:13}
\end{equation}
where
\begin{eqnarray}
V_{0}(x)&=&37.53[(1+\delta)^{5/3}+(1-\delta)^{5/3}]\rho_{0}^{2/3}(x)\nonumber \\
&+&b_{1}\rho_{0}(x)+b_{2}\rho_{0}^{4/3}(x)+b_{3}\rho_{0}^{5/3}(x)\nonumber \\
&+&\delta^{2}[b_{4}\rho_{0}(x)+b_{5}\rho_{0}^{4/3}(x)+b_{6}\rho_{0}^{5/3}(x)]
\label{eq:14}
\end{eqnarray}
with
\begin{eqnarray}
b_{1}&=&-741.28, \;\;\; b_{2}=1179.89, \;\;\; b_{3}=-467.54,\nonumber \\
b_{4}&=&148.26, \;\;\;\;\;\; b_{5}=372.84, \;\;\;\; b_{6}=-769.57.
\label{eq:15}
\end{eqnarray}
In Eq.~(\ref{eq:13}) $V_{0}(x)$ is the energy per particle in
nuclear matter (in MeV) accounting for the neutron-proton
asymmetry, $V_{C}$ is the Coulomb energy of protons in a flucton,
\begin{equation}
V_{C}=\frac{3}{5} \frac{Z^{2}e^{2}}{x},
\label{eq:16}
\end{equation}
and $V_{CO}$ is the Coulomb exchange energy:
\begin{equation}
V_{CO}=0.7386 Ze^{2} (3Z/4\pi x^{3})^{1/3}.
\label{eq:17}
\end{equation}
Thus, using the Brueckner theory, the symmetry energy $s^{ANM}(x)$
and the related quantities for ANM with density $\rho_{0}(x)$ (the
coefficient $a_{4}$ in Eq.~(\ref{eq:1})) have the forms:
\begin{eqnarray}
s^{ANM}(x)&=&41.7\rho_{0}^{2/3}(x)+b_{4}\rho_{0}(x) \nonumber \\
&+&b_{5}\rho_{0}^{4/3}(x)+b_{6}\rho_{0}^{5/3}(x),
\label{eq:18}
\end{eqnarray}
\begin{eqnarray}
p_{0}^{ANM}(x)&=&27.8\rho_{0}^{5/3}(x)+b_{4}\rho_{0}^{2}(x) \nonumber \\
&+&\frac{4}{3}b_{5}\rho_{0}^{7/3}(x)+\frac{5}{3}b_{6}\rho_{0}^{8/3}(x),
\label{eq:19}
\end{eqnarray}
and
\begin{eqnarray}
\Delta
K^{ANM}(x)&=&-83.4\rho_{0}^{2/3}(x)+4b_{5}\rho_{0}^{4/3}(x)\nonumber \\
&+&10b_{6}\rho_{0}^{5/3}(x).
\label{eq:20}
\end{eqnarray}
In our method (see also \cite{Gaidarov2011}) Eqs.~(\ref{eq:18}),
(\ref{eq:19}), and (\ref{eq:20}) are used to calculate the
corresponding quantities in finite nuclei $s$, $p_{0}$, and
$\Delta K$ from Eqs.~(\ref{eq:10}), (\ref{eq:11}), and
(\ref{eq:12}), respectively. We note that in the limit case when
$\rho(r)=\rho_{0}\Theta (R-r)$ and $|f(x)|^{2}$ becomes a $\delta$
function [see Eq.~(\ref{eq:9})], Eq.~(\ref{eq:10}) reduces to
$s^{ANM}(\rho_{0})=a_{4}$.

In our work we use the proton and neutron densities obtained from
self-consistent deformed Hartree-Fock calculations with
density-dependent Skyrme interactions \cite{Vautherin73} and
pairing correlations. Pairing between like nucleons is included by
solving the BCS equations at each iteration with a fixed pairing
strength that reproduces the odd-even experimental mass
differences \cite{Audi2003}.

The spin-independent proton and neutron densities are given by
\cite{Sarriguren2007,Guerra91}
\begin{equation}
\rho({\vec R})=\rho (r,z)=\sum _{i} 2v_i^2\rho_i(r,z)\, ,
\label{eq:21}
\end{equation}
where $r$ and $z$ are the cylindrical coordinates of ${\vec R}$, $v_i^2$ are the
occupation probabilities resulting from the BCS equations and $\rho_i$ are the
single-particle densities
\begin{equation}
\rho_i({\vec R})=  \rho_i(r,z)=|\Phi^+_i(r,z)|^2+
|\Phi^-_i(r,z)|^2
\label{eq:22}
\end{equation}
with
\begin{eqnarray}
\Phi^\pm _i(r,z)&=&{1\over \sqrt{2\pi}}\nonumber \\
&\times & \sum_{\alpha}\, \delta_{\Sigma, \pm 1/2}\,
\delta_{\Lambda,\Lambda^\mp}\, C_\alpha ^i\, \psi_{n_r}^\Lambda
(r) \, \psi_{n_z}(z)
\label{eq:23}
\end{eqnarray}
and $\alpha=\{n_r,n_z,\Lambda,\Sigma\}$. In (\ref{eq:23}) the
functions $\psi^{\Lambda}_{n_r}(r)$ and $\psi_{n_z}(z)$ are
expressed by Laguerre and Hermite polynomials:
\begin{equation}
\psi^{\Lambda}_{n_r}(r)=\sqrt{\frac{n_r}{(n_r+\Lambda )!}} \,
\beta_{\perp}\, \sqrt{2}\, \eta^{\Lambda/2}\, e^{-\eta/2}\,
L_{n_r}^{\Lambda}(\eta) \, ,
\label{eq:24}
\end{equation}
\begin{equation}
\psi_{n_z}(z)= \sqrt{\frac{1}{\sqrt{\pi}2^{n_z}n_z!}} \,
\beta^{1/2}_z\, e^{-{\xi}^2/2}\, H_{n_z}(\xi)
\label{eq:25}
\end{equation}
with
\begin{eqnarray}
\beta_z=(m\omega_z/\hbar )^{1/2}&,&\quad
\beta_\perp=(m\omega_\perp/\hbar )^{1/2},\nonumber \\
\quad \xi=z\beta_z&,&\quad \eta=r^2\beta_\perp ^2 \, .
\label{eq:26}
\end{eqnarray}
The normalization of the densities is:
\begin{equation}
\int \rho({\vec R}) d{\vec R} = X
\label{eq:27}
\end{equation}
with $X=Z,N$ for protons and neutrons, respectively.

The multipole decomposition of the density can be written in terms of even $\lambda$
multipole components as \cite{Vautherin73,Guerra91}
\begin{equation}
\rho(r,z)=\sum_{\lambda} \rho_{\lambda}(R)P_{\lambda}(\cos\theta).
\label{eq:27a}
\end{equation}
In the calculations, for the density distribution $\rho(r)$ needed to obtain
the weight function $|f(x)|^{2}$ [Eq.~(\ref{eq:9})] we use the monopole term
$\rho_{0}(R)$ in the expansion (\ref{eq:27a}).

The neutron skin thickness is usually estimated as the difference of the rms radii
of neutrons and protons:
\begin{equation}
\Delta R=<r_{\rm n}^2> ^{1/2}-<r_{\rm p}^2> ^{1/2}.
\label{eq:28}
\end{equation}
In our calculations the following Skyrme force parametrizations
are used: SLy4 \cite{sly4}, Sk3 \cite{sk3}, SGII \cite{sg2}, and
LNS \cite{Cao2006}. These are among the most extensively used
Skyrme forces that work successfully for describing finite nuclei
properties. Although it is well known that SGII and LNS
interactions do not predict accurate binding energies in finite
nuclei, we have included them in our work because they are
representative examples of Skyrme interactions and because we are
not concerned here with the absolute values of binding energies,
but rather with the isotopic evolution of relative differences of
various magnitudes and energy derivatives.

\section{Results and discussion}

\begin{figure*}
\centering
\includegraphics[width=140mm]{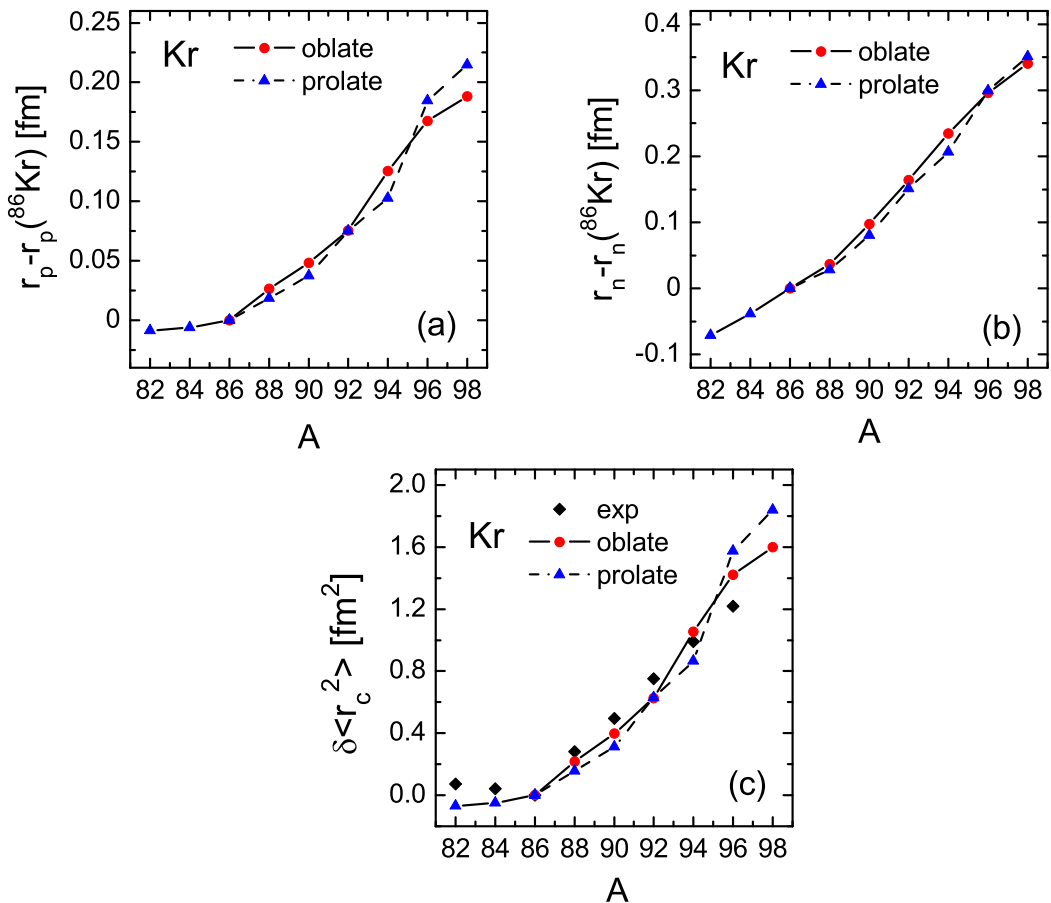}
\caption[]{(Color online) Proton (a) and neutron (b) rms radii
differences relative to $^{86}$Kr for oblate and prolate nuclear
shapes. (c) Theoretical (with SLy4 Skyrme force) and experimental
isotope shifts $\delta <r_{c}^{2}>$ of Kr isotopes relative to
$^{86}$Kr. The results for oblate and prolate shape for $A=82,84$
isotopes are indistinguishable.
\label{fig1}}
\end{figure*}

We start our analysis by searching for the role of deformation on
the rms radii. It is illustrated by showing the differences
between the proton (neutron) radii and the corresponding proton
(neutron) radius of semi-magic $^{86}$Kr ($N=50$) which is taken
as a reference nucleus. The results are given in Fig.~\ref{fig1}
[(a) and (b), respectively]. In addition, in Fig.~\ref{fig1}(c)
our results with SLy4 force for the squared charge radii
differences in Kr isotopes are shown and they are compared with
the experimental data from Ref.~\cite{Keim95}. Results of this
type have been presented in Ref.~\cite{Sarriguren2007}, where the
charge radii differences in Sn isotopes obtained from SLy4, SGII,
and Sk3 Skyrme forces are compared with the experimental data
taking the radius of $^{120}$Sn as the reference. In principle,
these differences have been found to be very sensitive probes of
nuclear shape transitions and it is worth studying globally a wide
region of neutron-rich exotic nuclei and discussing the
similarities and differences among the various isotopic chains
\cite{Guzman2011}.

In this analysis the even-even $^{86}$Kr isotope turns out to be
spherical, while for Kr isotopes beyond the semi-magic $^{86}$Kr
($N=50$) nucleus a deformation (oblate or prolate) takes place. We
note that for the considered lightest isotopes with $A=82,84$ both
shapes produce results that are indistinguishable and, therefore,
only the prolate solutions are presented. A smooth increase of
radii differences relative to $^{86}$Kr with increasing neutron
number can be observed in all three panels of Fig.~\ref{fig1}. The
charge and proton radii grow following the growth of the neutron
radius with increasing neutron number. These increases are only
slightly dependent on whether an oblate or a prolate shape is
considered. A similar behavior has been obtained from
Gogny-D1S-HFB calculations performed in Ref.~\cite{Guzman2011}. A
satisfactory agreement with the experimental isotope shifts is
observed in Fig.~\ref{fig1}(c) that provides a good starting point
to study further quantities such as the symmetry energy and
related characteristics of deformed nuclei within our theoretical
method.

\begin{figure*}
\centering
\includegraphics[width=110mm]{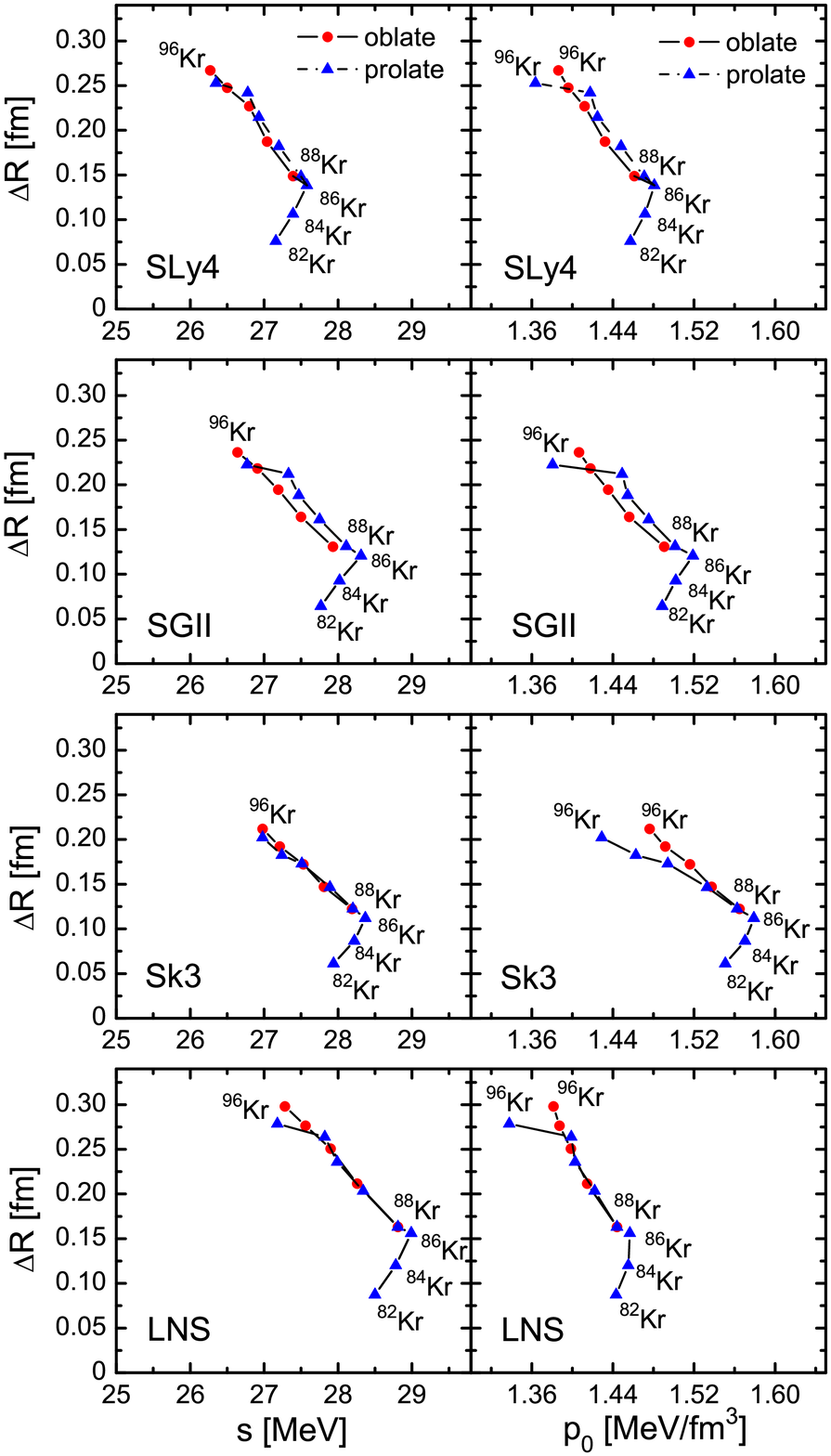}
\caption[]{(Color online) HF+BCS neutron skin thicknesses $\Delta
R$ for Kr isotopes as a function of the symmetry energy $s$ and
the pressure $p_{0}$ calculated with SLy4, SGII, Sk3, and LNS
forces and for oblate and prolate shapes. The results for oblate
and prolate shape for $A=82,84$ isotopes are indistinguishable.
\label{fig2}}
\end{figure*}

Next, an illustration of a possible correlation of the
neutron-skin thickness $\Delta R$ with the s and $p_{0}$
parameters extracted from the density dependence of the symmetry
energy around the saturation density for the Kr isotopic chain is
given in Fig.~\ref{fig2}. The symmetry energy and the pressure are
calculated within the CDFM according to Eqs.~(\ref{eq:10}) and
(\ref{eq:11}) by using the weight functions (\ref{eq:9})
calculated from the self-consistent densities in
Eq.~(\ref{eq:21}). The differences between the neutron and proton
rms radii of these isotopes [Eq.~(\ref{eq:28})] are obtained from
HF+BCS calculations using four different Skyrme forces, SLy4,
SGII, Sk3, and LNS. It can be seen from Fig.~\ref{fig2} that there
exists an approximate linear correlation between $\Delta R$ and
$s$ for the even-even Kr isotopes with $A=82-96$. Similarly to the
behavior of $\Delta R$ vs $s$ dependence for the cases of Ni and
Sn isotopes \cite{Gaidarov2011}, we observe a smooth growth of the
symmetry energy up to the semi-magic nucleus $^{86}$Kr ($N=50$)
and then a linear decrease of $s$ while the neutron-skin thickness
of the isotopes increases. This linear tendency expressed for Kr
isotopes with $A>86$ is similar for the cases of both oblate and
prolate deformed shapes. We note that all Skyrme parametrizations
used in the calculations reveal similar behavior; in particular,
the average slope of $\Delta R$ for various forces is almost the
same.

In addition, one can see from Fig.~\ref{fig2} a stronger deviation
between the results for oblate and prolate shape of Kr isotopes in
the case of SGII parametrization when displaying the correlation
between $\Delta R$ and $s$. This is valid also for the correlation
between $\Delta R$ and $p_{0}$, where more distinguishable results
for both types of deformation are present. The neutron skin
thickness $\Delta R$ for Kr isotopes correlates with $p_{0}$
almost linearly, as in the symmetry-energy case, with an inflexion
point transition at the semi-magic $^{86}$Kr nucleus. In addition,
one can see also from Fig.~\ref{fig2} that the calculated values
for $p_{0}$ are smaller in the case of LNS and SLy4 forces than
for the other two Skyrme parameter sets. In general, we would like
to note that the behavior of deformed Kr isotopes shown in
Fig.~\ref{fig2} is comparable with the one found for the spherical
Ni and Sn isotopes having a magic proton number that we discussed
in Ref.~\cite{Gaidarov2011}. The small differences just indicate
that stability patterns are less regular within isotopic chains
with a non-magic proton number.

\begin{figure}
\centering
\includegraphics[width=90mm]{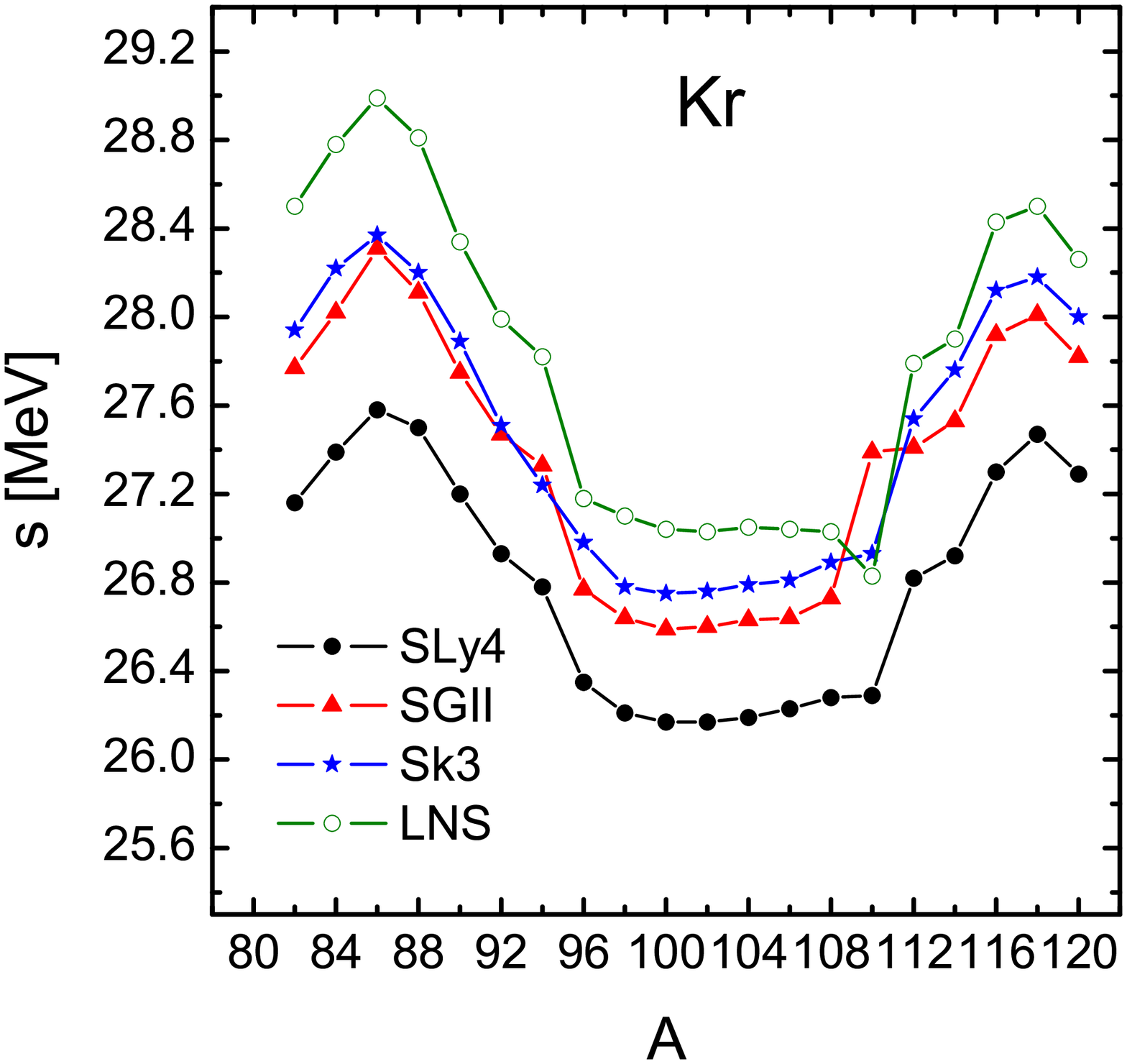}
\caption[]{(Color online) The symmetry energies $s$ for Kr
isotopes ($A=82-120$) calculated with SLy4, SGII, Sk3, and LNS
forces. \label{fig3}}
\end{figure}

For more complete study, we also consider in our work the
extremely neutron-rich Kr isotopes ($A=96-120$). The results for
the symmetry energy $s$ as a function of the mass number $A$ for
the whole Kr isotopic chain ($A=82-120$) are presented in
Fig.~\ref{fig3}. We observe peaks of the symmetry energy at
specific Kr isotopes, namely at semi-magic $^{86}$Kr ($N=50$) and
$^{118}$Kr ($N=82$) nuclei. In addition, a flat area is found
surrounded by transitional regions $A=88-96$ and $A=110-116$.
Also, the SGII and Sk3 forces yield values of $s$ comparable with
each other that lie between the corresponding symmetry energy
values when using SLy4 and LNS sets. The specific nature of LNS
force \cite{Cao2006} (not being fitted to finite nuclei) leads to
larger values of $s$ (and to a larger size of the neutron-skin
thickness, as it is seen from Fig.~\ref{fig2}) with respect to the
results with other three forces. Although the values of $s$
slightly vary within the Kr isotopic chain when using different
Skyrme forces, the curves presented in Fig.~\ref{fig3} exhibit the
same trend.

\begin{figure}
\centering
\includegraphics[width=80mm]{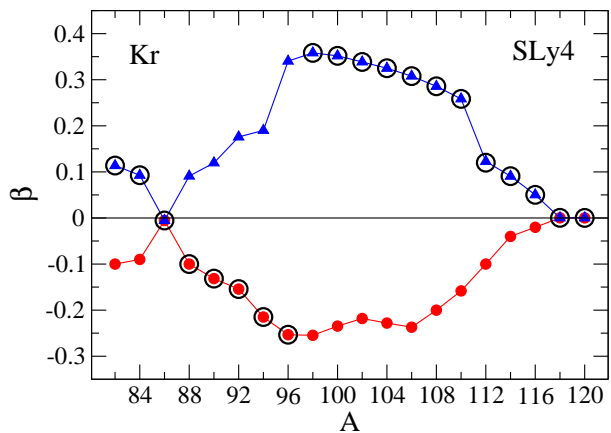}
\caption[]{(Color online) The quadrupole parameter $\beta$ as a
function of the mass number $A$ for the even-even Kr isotopes
($A=82-120$) in the case of SLy4 force.
\label{fig4}}
\end{figure}

The results shown in Fig.~\ref{fig3} are closely related to the
evolution of the quadrupole parameter
$\beta=\sqrt{\pi/5}Q/(A\langle r^2\rangle ^{1/2})$ ($Q$ being the
mass quadrupole moment and $\langle r^2\rangle ^{1/2}$ the nucleus
rms radius) as a function of the mass number $A$ that is presented
in Fig.~\ref{fig4}. First, one can see from Fig.~\ref{fig4} that
the semi-magic $A=86$ and $A=118$ Kr isotopes are spherical, while
the open-shell Kr isotopes within this chain possess two
equilibrium shapes, oblate and prolate. In the case of open-shell
isotopes, the oblate and prolate minima are very close in energy
and the energy difference is always less than 1 MeV. In this
region of even-even Kr isotopes with very large $N/Z$ ratio ($\geq
1.7$) the competition between the prolate and oblate shapes has
also been studied with HFB calculations and the Gogny force in
Ref.~\cite{Hilaire2007}. Shape coexistence in lighter Kr isotopes
has also been examined \cite{Sarriguren99,Flocard73}.
Nevertheless, we specify in Fig.~\ref{fig4} which shape
corresponds to the ground state of each isotope by encircling
them. Thus, the trend that the evolution of the symmetry energy
shown in Fig.~\ref{fig3} follows can be clearly understood.  The
peaks of the symmetry energy correspond to the closed-shell nuclei
that are spherical. Mid-shell nuclei ($A=96-110$) are well
deformed and exhibit a stabilized behavior with small values of
$s$. The transitional regions from spherical to well deformed
shapes correspond to transitions from the peaks to the valley in
the symmetry energy.

\begin{figure*}
\centering
\includegraphics[width=170mm]{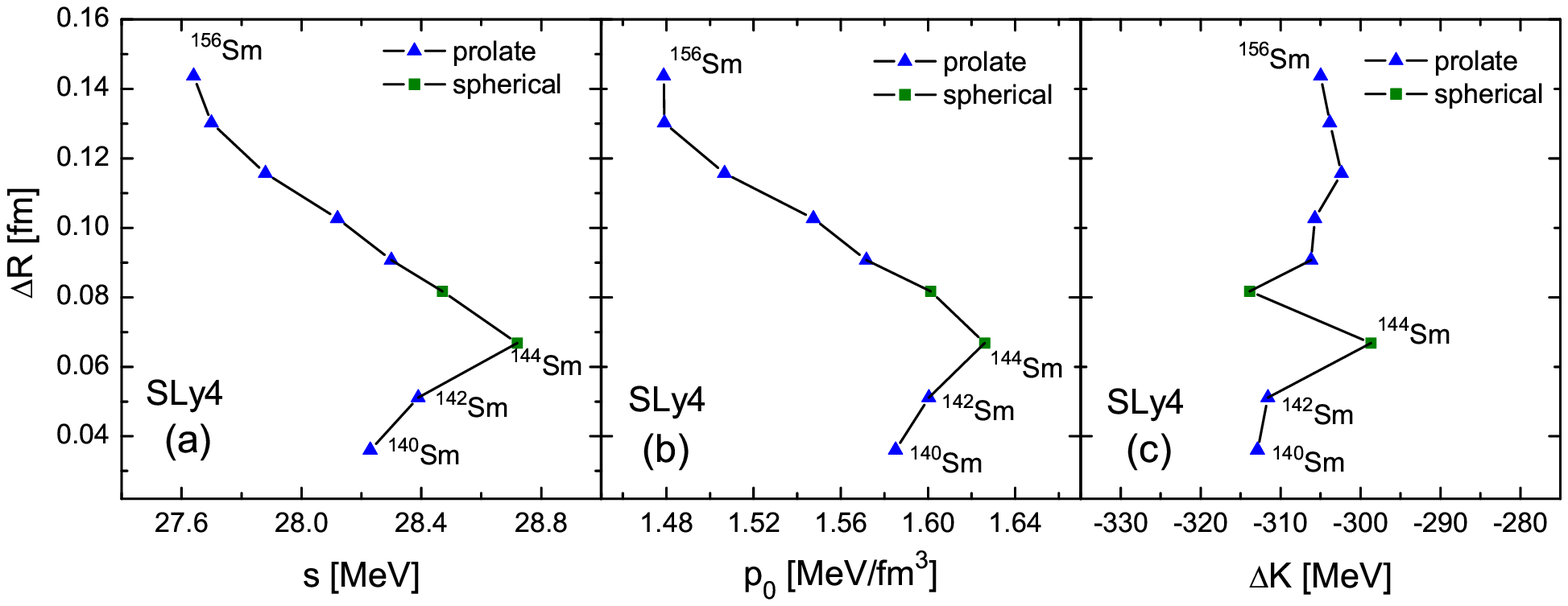}
\caption[]{(Color online) HF+BCS neutron skin thicknesses $\Delta
R$ for Sm isotopes as a function of the symmetry energy $s$ (a),
pressure $p_{0}$ (b), and asymmetric compressibility $\Delta K$
(c) calculated with SLy4 force.
\label{fig5}}
\end{figure*}

\begin{figure*}
\centering
\includegraphics[width=150mm]{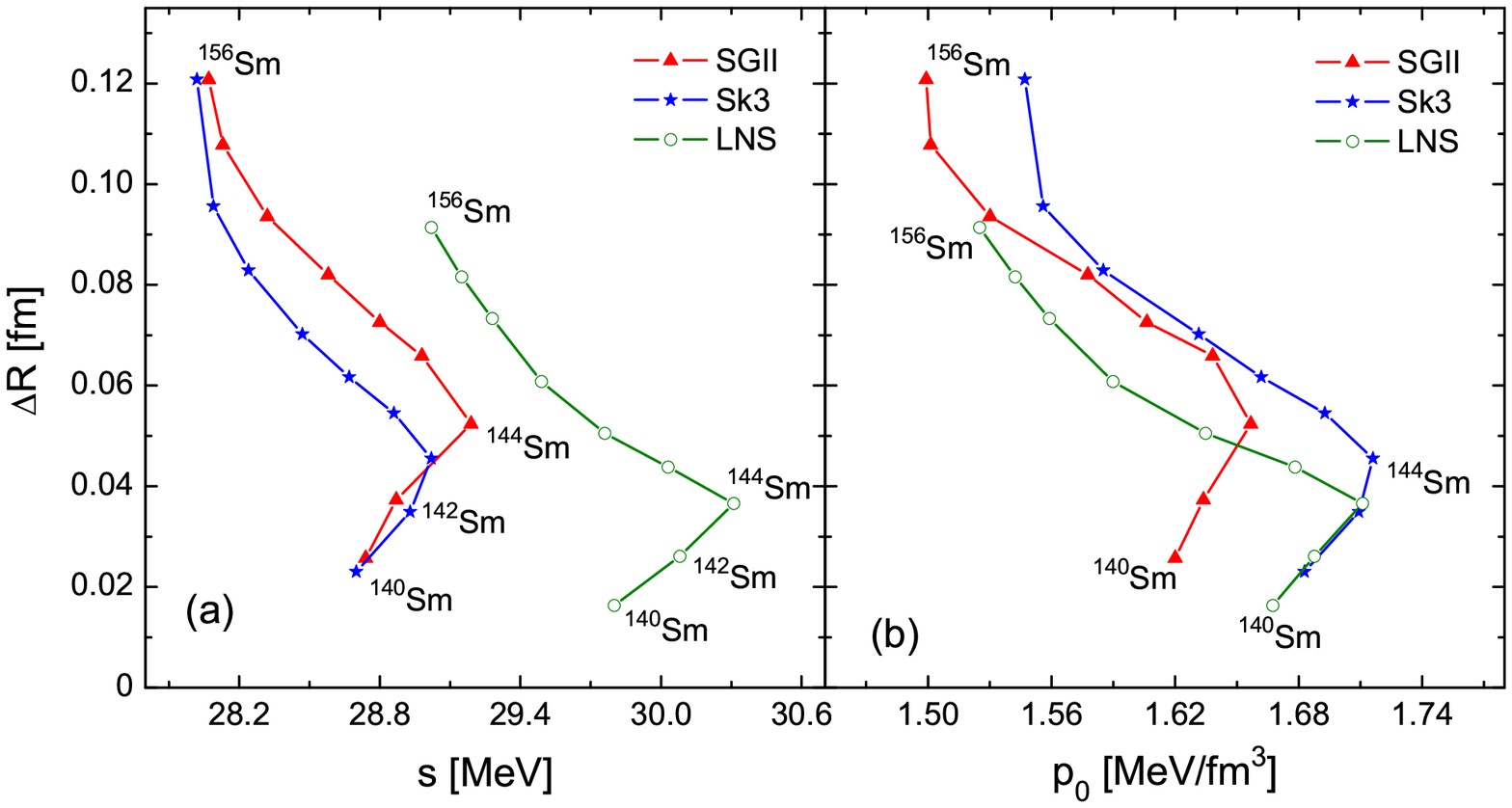}
\caption[]{(Color online) HF+BCS neutron skin thicknesses $\Delta
R$ for Sm isotopes as a function of the symmetry energy $s$ (a)
and the pressure $p_{0}$ (b) calculated with SGII, Sk3, and LNS
forces. \label{fig6}}
\end{figure*}

In Figs.~\ref{fig5} and \ref{fig6} we give results for Sm isotopes
($A=140-156$) as a well established example of deformed nuclei. In
the calculations, all Sm isotopes are found to have a prolate
shape, except for the even-even $^{144}$Sm and $^{146}$Sm nuclei
that are spherical. Such an evolution of shape from the spherical
to the axially deformed shapes in the same Sm isotopic region is
in accordance with the results obtained from microscopic
calculations in the RMF theory \cite{Meng2005}. In
Ref.~\cite{Meng2005} the ground state of the semi-magic $^{144}$Sm
($N=82$) is found to be spherical (having about 12 MeV stiff
barrier against deformation) and the deformation in $^{146}$Sm to
be still small. With the increase of the neutron number, the
ground state gradually moves toward the deformed one till the well
deformed $^{154-158}$Sm \cite{Meng2005}. Also, the analysis of the
potential energy curves \cite{Guzman2007} within the same
microscopic approach that we use in the present work, as well as
within the HFB method with Gogny interaction \cite{Robledo2008},
confirmed the transitional behavior between the spherical
$^{144}$Sm and the well prolate deformed $^{154-158}$Sm isotopes.

The results for the correlation between the neutron skin thickness
and the nuclear matter properties in finite nuclei with SLy4
Skyrme force for a chain of Sm isotopes are shown in
Fig.~\ref{fig5}, while those with SGII, Sk3, and LNS forces are
presented in Fig.~\ref{fig6}. Similar to the case of Kr isotopes
with transition at specific shell closure, we observe a smooth
growth of the symmetry energy until the semi-magic nucleus
$^{132}$Sm ($N=82$) and then an almost linear decrease of $s$
while the neutron skin thickness of the isotopes increases. An
approximate linear correlation between $\Delta R$ and $p_{0}$ is
also shown in Figs.~\ref{fig5}(b) and \ref{fig6}(b), while
Fig.~\ref{fig5}(c) exhibits a very irregular behavior of $\Delta
R$ as a function of the asymmetric compressibility $\Delta K$.
Nevertheless, the values of $\Delta K$ deduced from our
calculations are in the interval between --295 and --315 MeV that
compare fairly well with the neutron-asymmetry compressibility
($K_{\Sigma}^{\prime}=-320\pm 180$ MeV) deduced from the data
\cite{Sharma88} on the breathing mode giant monopole resonances in
the isotopic chains of Sm and Sn nuclei.

\begin{figure*}
\centering
\includegraphics[width=170mm]{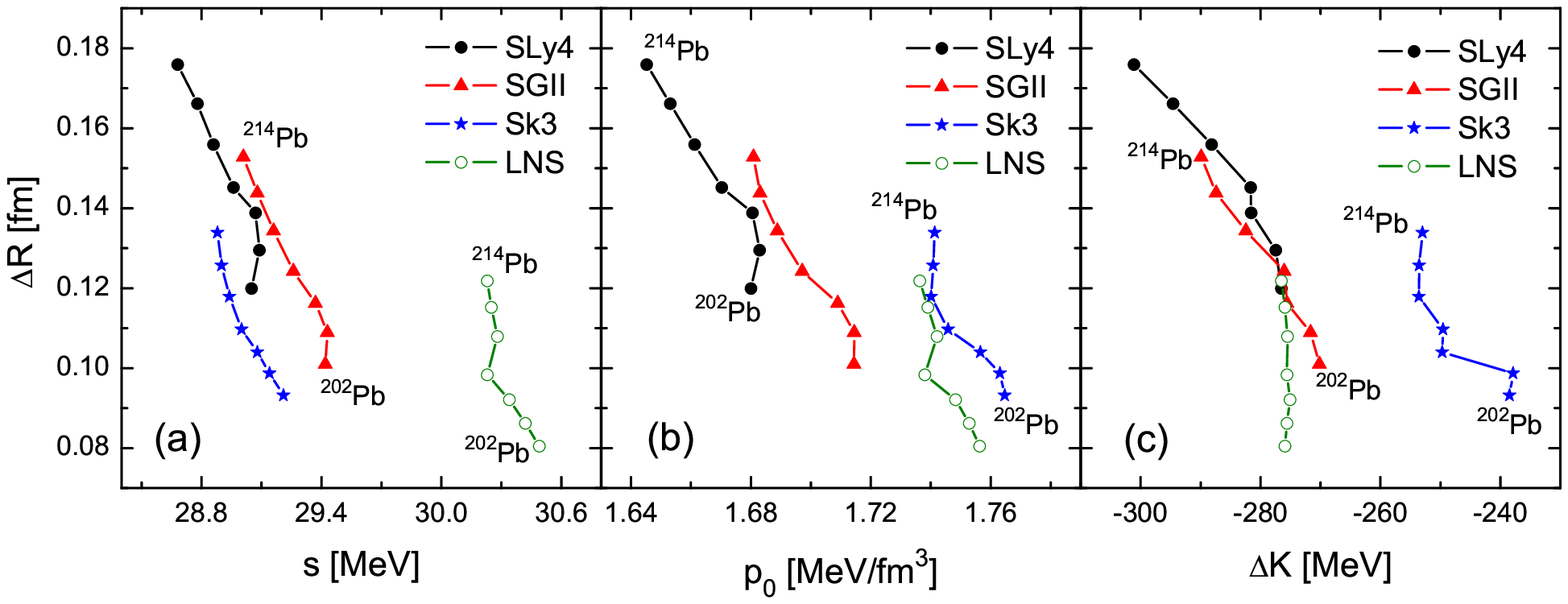}
\caption[]{(Color online) HF+BCS neutron skin thicknesses $\Delta
R$ for Pb isotopes as a function of the symmetry energy $s$ (a),
pressure $p_{0}$ (b), and asymmetric compressibility $\Delta K$
(c) calculated with SLy4, SGII, Sk3, and LNS force. \label{fig7}}
\end{figure*}

The theoretical neutron skin thickness $\Delta R$ of Pb nuclei
($A=202-214$) against the parameters of interest, $s$, $p_{0}$,
and $\Delta K$, is illustrated in Fig.~\ref{fig7}. In this work we
consider an extended chain of Pb isotopes in comparison to the one
analyzed in Ref.~\cite{Gaidarov2011} by adding two nuclei lighter
than $^{206}$Pb. Therefore, a more precise study of the
corresponding correlations, especially in the transition region at
the double-magic $^{208}$Pb nucleus, could be made. Here, we test
three more parametrizatios (SGII, Sk3, and LNS) in addition to the
SLy4 force applied in \cite{Gaidarov2011}. All predicted
correlations manifest an almost linear dependence and no
pronounced kink at $^{208}$Pb is observed. Similarly to Kr and Sm
isotopes presented in this study (and isotopes from Ni and Sn
chains described in \cite{Gaidarov2011}), the LNS force produces
larger symmetry energies $s$ than the other three forces also for
Pb nuclei with values exceeding 30 MeV. Another peculiarity of the
results obtained with LNS is the almost constant $\Delta K$
observed in Fig.~\ref{fig7}a.

Further attention deserves the value of the neutron skin thickness
in $^{208}$Pb, whose determination has motivated recent
experiments. The model-independent measurement of parity-violating
asymmetry (which is sensitive to the neutron distribution) in the
elastic scattering of polarized electrons from $^{208}$Pb at JLAB
within the PREX Collaboration \cite{prex,Abrahamyan2012} has
provided the first electroweak observation of $\Delta
R$=$0.33^{+0.16}_{-0.18}$ fm in $^{208}$Pb. Obviously, future
precise measurements are needed to reduce the quoted uncertainties
of $\Delta R$. The distorted wave electron scattering calculations
for $^{208}$Pb \cite{Horowitz2012} extracted a result for the
neutron skin thickness which agrees with that reported in the
experimental paper \cite{Abrahamyan2012}. As can be seen from
Fig.~\ref{fig7}(a), the value of $\Delta R$ for $^{208}$Pb (0.1452
fm) deduced from the present HF+BCS calculations with SLy4 force
agrees with the recent experimentally extracted skin thickness
($0.156^{+0.025}_{-0.021}$ fm) using its correlation with the
dipole polarizability \cite{Tamii2011}. However, this experimental
value was derived by means of covariance analysis based on one
Skyrme functional (SV-min). In this respect, a systematic study
with a variety of EDFs as well as experimental tests in other
nuclei would be important because the correlation between
polarizability, neutron skin thickness, and symmetry energy is
model-dependent (see, for example, Fig.~1 of
Ref.~\cite{Piekarewicz2012}). In addition, our theoretically
obtained value of $\Delta R$ for $^{208}$Pb agrees well with the
value $0.18\pm 0.027$ fm from Ref.~\cite{Tsang2012}. It is lower
than the one obtained in Refs.~\cite{Chen2005} and
\cite{Agrawal2010} with the same Skyrme force, but is in agreement
with the values calculated with self-consistent densities of
several nuclear mean-field models (see Table I in
Ref.~\cite{Centelles2010}). The $p_{0}$ and $\Delta K$ values for
$^{208}$Pb are in a good agreement with those from
Ref.~\cite{Chen2005}.

We would like to note that linear correlations $\Delta R$ vs. $s$
and $\Delta R$ vs. $p_{0}$ are found to exist in the Kr
($A=82-96$), Sm ($A=140-156$) and Pb ($A=202-214$) isotopic
chains. The correlation $\Delta R$ vs. $\Delta K$ presented on the
examples of Sm and Pb isotopes (Figs.~\ref{fig5} and \ref{fig7},
respectively) is less strong than the ones mentioned before and
shows a slight irregular behavior. This concerns also the
calculated results for $\Delta K$ in the case of Kr isotopes that
are not shown in Fig.~\ref{fig2}. Such observations are confirmed
by the results for these correlations obtained in
Refs.~\cite{Vidana2009,Yoshida2004}. In addition, it was
demonstrated in \cite{Reinhard2010} that among various observables
that correlate with the neutron form factor (related to the
neutron density and, thus, to the neutron skin thickness) the
incompressibility is a poor indicator of isovector properties.

Following our previous analysis within the CDFM approach
\cite{Gaidarov2011}, here we would like to give more detailed
study of the weight function $|f(x)|^{2}$ (that is related to the
density and thus, to the structural peculiarities) to understand
the kinks observed in the relationships between $\Delta R$ and
$s$, as well as $\Delta R$ and $p_{0}$. The latter were shown to
exist \cite{Gaidarov2011} in double-magic nuclei in the cases of
Ni (at $^{78}$Ni) and Sn (at $^{132}$Sn) isotopic chains. As one
can see in Figs.~\ref{fig2}, \ref{fig5}, and \ref{fig6} of the
present work, they exist also in the considered cases of Kr (at
$^{86}$Kr) and Sm (at $^{144}$Sm) isotopes. In contrast, such a
kink does not exist in the case of Pb isotopic chain (at
$^{208}$Pb, particularly).

\begin{figure*}
\centering
\includegraphics[width=175mm]{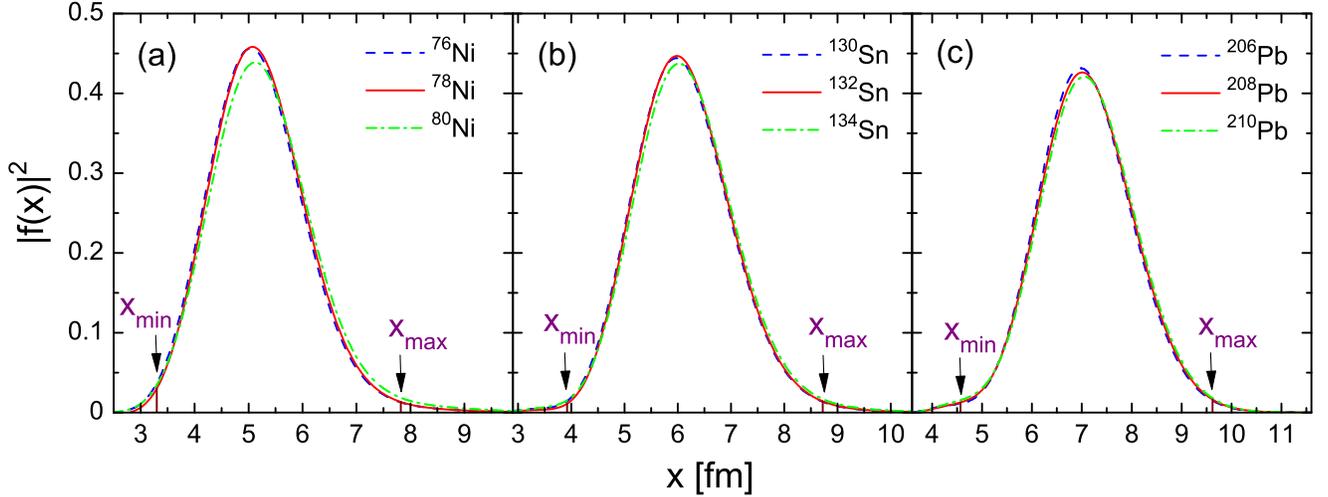}
\caption[]{(Color online) The weight function $|f(x)|^{2}$
calculated for three Ni ($A=76, 78, 80$) isotopes (a), three Sn
($A=130, 132, 134$) isotopes (b), and three Pb ($A=206, 208, 210$)
isotopes (c) by using the HF+BCS total densities for these nuclei
and with SLy4 force.
\label{fig8}}
\end{figure*}

\begin{figure}
\centering
\includegraphics[width=80mm]{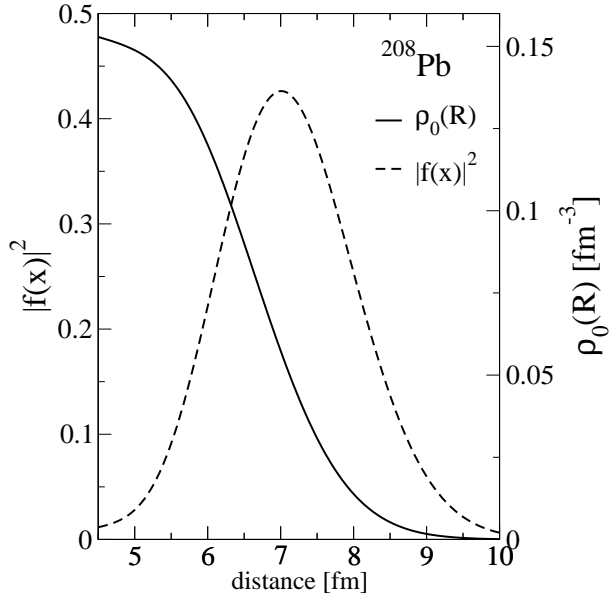}
\caption[]{The weight function $|f(x)|^{2}$ and the HF+BCS total
density $\rho_{0}(R)$ of $^{208}$Pb calculated with SLy4 force.
\label{fig9}}
\end{figure}

Here we analyze, as an example, the cases of Ni, Sn and Pb
isotopic chains, trying to understand the origin of the kinks
without additional complexities coming from deformation. For this
purpose, let us introduce the quantity
\begin{equation}
\Delta s_{\pm}=\frac{s_{A\pm 2}-s_{A}}{s_{A}}
\label{eq:29}
\end{equation}
which gives information on the relative deviation of the symmetry
energy $s$ of even-even isotopes with respect to the double-magic
ones, namely with $A=78$ for Ni, $A=132$ for Sn, and $A=208$ for
Pb. Here we will consider the range of integration on $x$ in
Eq.~(\ref{eq:10}) in each of the cases of Ni, Sn, and Pb isotopes.
Firstly, we will introduce the value of $x_{\text{min}}$ at which
the symmetry energy for nuclear matter $s^{ANM}(x)$ changes sign
from negative (at $x<x_{\text{min}}$) to positive (at
$x>x_{\text{min}}$). In our work we use the symmetry energy
$s^{ANM}(\rho(x))$ [Eq.~(\ref{eq:18})] from the Brueckner theory,
that is defined by the second derivative of the energy per
particle $E(\rho,\delta)$. Considering in principle in the CDFM
the range of $x$ from zero to infinity we include in this way the
region of densities $\rho_{0}(x)$ from infinity to zero,
respectively. When the values of $x$ are small we consider in
practice values of the density $\rho_{0}(x)$ that are much larger
than the density in the equilibrium state $\rho_{0}$. In this case
unphysical (negative) values of the symmetry energy appear, and
thus starting the integration from $x\geq x_{\text{min}}$ we
exclude these values. We have to note simultaneously that at
$x<x_{\text{min}}$ the weight function $|f(x)|^{2}$ is close to
zero (it is its "left wing", see Fig.~\ref{fig8}), so there is no
contribution to $s$ from this region ($x<x_{\text{min}}$).
Secondly, we introduce the value of $x_{\text{max}}$ (in the
"right wing" of $|f(x)|^{2}$) beyond which the contribution to $s$
(i.e. the result of the integration in Eq.~(\ref{eq:10}) from
$x_{\text{max}}$ to infinity) is negligible. If we define by
$\Delta x=x_{\text{max}}-x_{\text{min}}$, then we impose for
$x_{\text{max}}$ the condition $s-s_{\Delta x}\leq 0.1$ MeV, where
$s_{\Delta x}$ is obtained by Eq.~(\ref{eq:10}) integrating over
$x$ from $x_{\text{min}}$ to $x_{\text{max}}$, while $s$ is the
result of integration from $x_{\text{min}}$ to infinity. The
points $x_{\text{min}}$ and $x_{\text{max}}$ corresponding to the
double-magic nuclei of the three isotopic chains considered are
indicated in Fig.~\ref{fig8}, where the intrinsic weight functions
$|f(x)|^{2}$ for $^{76,78,80}$Ni [Fig.~\ref{fig8}(a)],
$^{130,132,134}$Sn [Fig.~\ref{fig8}(b)], and $^{206,208,210}$Pb
[Fig.~\ref{fig8}(c)], respectively, are separately presented.
Then, if one uses Eq.~(\ref{eq:10}) that determines the symmetry
energy $s$ in finite nuclei within the CDFM, the contribution
$s_{\Delta x}$ of $s$ in the interval $\Delta x$ is obtained. The
values of $s_{\Delta x}$ are given in Table~\ref{table1} together
with the values of $x_{\text{min}}$, $x_{\text{max}}$, and $s$. We
can see from Table~\ref{table1} that, as can be expected, the
total symmetry energy is almost exhausted by its contribution
$s_{\Delta x}$, thus showing the important role which the nuclear
surface plays for the obtained symmetry energy values. For a
better illustration of this fact, we plot together in
Fig.~\ref{fig9} the weight function $|f(x)|^{2}$ and the HF+BCS
total density $\rho_{0}(R)$ of $^{208}$Pb in the case of SLy4
force. As can be seen, the nuclear surface part of the density
which determines the weight function $|f(x)|^{2}$ containing the
peak around the maximum is responsible to a large extent for the
main contribution to the total symmetry energy.

\begin{table*}
\caption{Integration limits values $x_{\text{min}}$ and
$x_{\text{max}}$ (in fm), values of the contribution $s_{\Delta
x}$ to the total symmetry energy $s$ (in MeV) for three Ni ($A=76,
78, 80$), Sn ($A=130, 132, 134$), and Pb ($A=206, 208, 210$)
isotopes.}
\begin{center}
{\begin{tabular}{ccccccccccccccccccccccc}
\hline\noalign{\smallskip} & & & $^{76}$Ni & & $^{78}$Ni & &
$^{80}$Ni & & & $^{130}$Sn & & $^{132}$Sn & & $^{134}$Sn & & &
$^{206}$Pb & & $^{208}$Pb & & $^{210}$Pb\\
\noalign{\smallskip}\hline\noalign{\smallskip}
$x_{\text{min}}$  & & & 3.26  & &  3.30  & & 3.32  & & & 3.90  & & 3.92  & & 3.94  & & & 4.56  & &  4.56  & & 4.58\\
$x_{\text{max}}$  & & & 7.78  & &  7.82  & & 8.16  & & & 8.70  & & 8.72  & & 8.92  & & & 9.56  & &  9.62  & & 9.64\\
$s_{\Delta x}$    & & & 27.55 & &  27.75 & & 27.37 & & & 28.52 & & 28.66 & & 28.46 & & & 28.97 & &  28.86 & & 28.76\\
$s$               & & & 27.65 & &  27.85 & & 27.47 & & & 28.62 & & 28.76 & & 28.56 & & & 29.07 & &  28.96 & & 28.86\\
\noalign{\smallskip} \hline
\end{tabular}}
\end{center}
\label{table1}
\end{table*}

The analysis of the quantity $\Delta s_{\pm}$ [{Eq.~(\ref{eq:29})]
might be instructive because this quantity is a direct measure of
the relative deviation of the symmetry energy with respect to the
double-magic nuclei taking them as reference nuclei in each of the
chains, where the kinks are expected. The values of $\Delta s_{+}$
and $\Delta s_{-}$ are listed in Table~\ref{table2}, where the two
numbers for each isotopic chain correspond to the range of
integration $\Delta x$. One can see firstly from this Table that
the absolute values of $\Delta s_{+}$ and $\Delta s_{-}$ for Pb
isotopes are comparable with each other, which is not the case for
the two other isotopic chains. Secondly, and very important is
that the $\Delta s_{+}$ value turns out to be negative and $\Delta
s_{-}$ value to be positive for Pb isotopes at the range of
integration $\Delta x$, and this is the main difference regarding
to the corresponding values (both are negative) in the Ni and Sn
chains.

\begin{table}
\caption{Relative deviation values of the symmetry energy $\Delta
s_{+}$ and $\Delta s_{-}$ [Eq.~(\ref{eq:29})] for the range of
integration $\Delta x$ in Eq.~(\ref{eq:10}) and for Ni, Sn, and Pb
isotopes.}
\begin{center}
{\begin{tabular}{cccccccccccc}
\hline\noalign{\smallskip} & & & Ni & & & Sn & & & Pb \\
\noalign{\smallskip}\hline\noalign{\smallskip}
$\Delta s_{+}$  & & &  -0.0137 & & &  -0.0070 & & & -0.0035 \\
$\Delta s_{-}$  & & &  -0.0072 & & &  -0.0049 & & &  0.0038 \\
\noalign{\smallskip} \hline
\end{tabular}}
\end{center}
\label{table2}
\end{table}

These differences can be attributed to the profiles of the density
distributions, particularly in the surface region. They are given
in Fig.~\ref{fig10}, where curves for five Ni, Sn, Pb isotopes
around double-magic $^{78}$Ni, $^{132}$Sn, and $^{208}$Pb nuclei
are presented in panels (a), (b), and (c), respectively. One can
see from Fig.~\ref{fig10} the same trend in the tails of the three
isotopic chains, which are ordered according to the mass number
$A$, being higher for heavier isotopes to produce larger radii. On
the other hand the behavior in the top part of the surface region,
shown in the inset of the panels, is different. In the case of Ni
and Sn isotopes in panels (a) and (b), one observes that the
double-magic nuclei have the largest density with all the
neighboring isotopes lying below. In the case of Pb isotopes in
panel (c), the density increases from heavier to lighter isotopes
with the double-magic nucleus in between. In Pb isotopes, this
ordering is opposite in the tail. As a result of this, the slope
of the density in Pb isotopes, and therefore $|f(x)|^{2}$,
decreases with the number of neutrons continuously and no kink is
present in the symmetry energy. On the other hand, in the case of
Ni and Sn isotopes, the slope of the density is larger for the
double-magic isotopes generating a kink in the symmetry energy.

\begin{figure*}
\centering
\includegraphics[width=160mm]{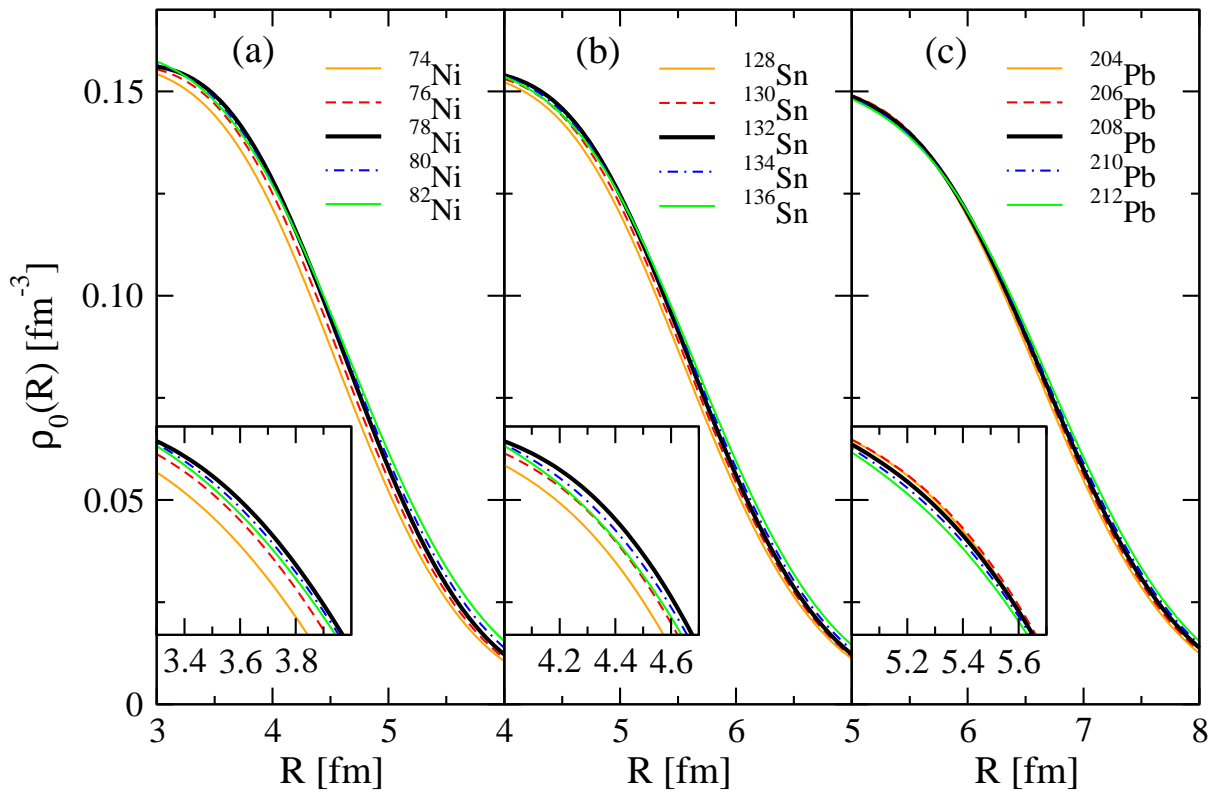}
\caption[]{(Color online) HF+BCS total densities in the surface
region for five Ni ($A$=74--82) isotopes (a), five Sn
($A$=128--136) isotopes (b), and five Pb ($A$=204--212) isotopes
(c) around double-magic $^{78}$Ni, $^{132}$Sn, and $^{208}$Pb
nuclei, respectively, calculated with SLy4 force.
\label{fig10}}
\end{figure*}

In this way, the kinks displayed in our previous study
\cite{Gaidarov2011} by the Ni and Sn isotopes and the fact that no
kinks appear in the Pb chain considered can be understood. It
concerns not only the symmetry energy evolution, but also its
relationship with the neutron skin thickness $\Delta R$ and the
pressure $p_{0}$. We would like to summarize the discussion about
the kinks, saying that within the CDFM we are able to use only the
densities resulting from the specific shells that are occupied.
The presented Figures \ref{fig9} and \ref{fig10} give information
about the differences between densities of Ni, Sn, and Pb isotopes
that are reflected in the corresponding weight functions
$|f(x)|^{2}$ for these isotopes (see Fig.~\ref{fig8}). Of course,
these differences are due to the different occupancies of the
shells in these nuclei. Thus, we see that in general kinks are
produced at shell closures, but the analyses of the precise
dependence of various kinks on the amount of occupation of
specific shells will require further work.

\section{Conclusions}

In this paper, we have investigated possible relationships between
the neutron skin thickness of deformed neutron-rich nuclei and the
symmetry energy characteristics of nuclear matter for these
nuclei. A microscopic approach based on deformed HF+BCS
calculations with Skyrme forces has been used. Four Skyrme
parametrizations were involved in the calculations: SGII, Sk3,
SLy4, and LNS. Nuclear matter properties of nuclei from Kr and Sm
isotopic chains have been studied by applying the CDFM that
provides a transparent and analytic way to calculate the intrinsic
EOS quantities by means of a convenient approach to the weight
function. As a first step we study the variation with neutron
number of proton and neutron radii predicted by the
self-consistent microscopic (HF+BCS) calculations. We find that
charge and neutron radii increase similarly with neutron number
following the increase of neutron radii, as it should be expected
from the general properties of the nuclear force. The same
microscopic method is then applied to calculate the skin thickness
and the weight function for each isotope. The analysis of the
nuclear symmetry energy $s$, the neutron pressure $p_{0}$, and the
asymmetric compressibility $\Delta K$ has been carried out on the
basis of the Brueckner EDF for infinite nuclear matter.

For both Kr ($A=82-96$) and Sm ($A=140-156$) isotopic chains we
have found that there exists an approximate linear correlation
between the neutron skin thickness of these nuclei and their
nuclear symmetry energies. Comparing with the spherical case of
Ni, Sn, and Pb nuclei described in our previous study
\cite{Gaidarov2011}, we note that the linear correlation observed
in the Kr and Sm isotopes is not smooth enough due to their
different equilibrium shapes, as well as to the transition regions
between them. As known, the latter are difficult to be interpreted
as they exhibit a complicated interplay of competing degrees of
freedom. Nevertheless, a smoother behavior is observed in Kr
isotopes that is a consequence of the stabilization of the oblate
shapes along the isotopic chain. As far as Sm isotopes are
concerned, the shape evolution from the spherical to the axially
deformed configurations in the Sm isotopes causes a less
pronounced linearity of the observed correlation between $\Delta
R$ and $s$. A similar correlation between $\Delta R$ and $p_{0}$
is also found to exist, while the relation between $\Delta R$ and
$\Delta K$ exhibits an irregular behavior. However, for both
classes of deformed nuclei an inflection point transition at
specific shell closure, in particular at semi-magic $^{86}$Kr and
$^{144}$Sm nuclei, appears for these correlations of the neutron
skins with $s$ and $p_{0}$. In addition, the role of the relative
neutron-proton asymmetry on the evolution of the symmetry energy
has been pointed out on the example of Kr isotopes with $N>60$.

We have analyzed in detail the existence of kinks on the example
of the Ni and Sn isotopic chains and the lack of such kink for the
Pb isotopic chain. For this purpose, we have estimated the
relative deviation of the symmetry energy of even-even isotopes
with respect to the double-magic $^{78}$Ni, $^{132}$Sn, and
$^{208}$Pb nuclei. An analysis of the weight function $|f(x)|^{2}$
(determined by the HF+BCS density distributions) as the key
ingredient of the CDFM is performed. It is shown that for the Pb
isotopes the different signs of the relative deviations
corresponding to the range of integration on $x$ in
Eq.~(\ref{eq:10}) that contains the peak of $|f(x)|^{2}$ is in
favor of the absence of kink in the Pb chain. Thus, from the
previous study in Ref.~\cite{Gaidarov2011} and the present
analysis the kinks displayed by the Ni and Sn can be understood as
consequences of particular differences in the structure of these
nuclei and the resulting densities and weight functions.

It has to be mentioned that the used microscopic theoretical
approach is capable also to predict important nuclear matter
quantities in deformed neutron-rich exotic nuclei and their
relation to surface properties of these nuclei. This is confirmed
by the good agreement achieved with other theoretical predictions
and some experimentally extracted ground-state properties. New
experimental results on giant resonances, neutron skin in heavy
nuclei, and heavy-ion collisions could further lead to new
constraints, e.g. on the nuclear symmetry energy, allowing
successful interpretation of data and observations on physical
quantities of nuclear systems.

\begin{acknowledgments}
Two of the authors (M.K.G. and A.N.A.) are grateful for the
support of the Bulgarian Science Fund under Contract No.~02--285.
E.M.G. and P.S. acknowledge support from MINECO (Spain) under
Contracts FIS2011--23565 and FPA2010--17142.
\end{acknowledgments}

\end{document}